\documentclass[12pt,tightenlines,eqsecnum,floats,showpacs,nofootinbib,amsmath,amssymb,aps]{revtex4}

\usepackage{amsmath,amssymb,amsfonts,amsthm,amscd}
\usepackage{graphicx}
\usepackage{enumerate}
\usepackage{colordvi}
\usepackage{epsfig}
\usepackage{natbib}
\usepackage{enumerate}
\usepackage{color}
\def\be{\begin{equation}}
\def\ee{\end{equation}}
\def\ba{\begin{eqnarray}}
\def\ea{\end{eqnarray}}

\def\Lie{\mathcal{L}}
\def\f{\frac}

\def\H{\mathcal{H}}
\def\d{{\rm{d}}}

\begin{document}

\title{Inflationary Attractors and their Measures}
\author{Alejandro Corichi$^1$}
\email{corichi@matmor.unam.mx}
\author{David Sloan$^2$}
\email{djs228@hermes.cam.ac.uk}
\affiliation{$^1$ Centro de Ciencias Matem\'{a}ticas, Universidad Nacional Aut\'{o}noma de M\'{e}xico, UNAM-Campus Morelia, A. Postal 61-3, Morelia, Michoac\'{a}n 58090, Mexico \\
$^{2}$ DAMTP, Centre for Mathematical Sciences, Wilberforce Rd., Cambridge
University, Cambridge CB3 0WA, UK}

\begin{abstract}

Several recent misconceptions about the measure problem in inflation and the nature of inflationary attractors are addressed. We clarify some issues regarding the Hamiltonian dynamics of a flat Friedmann-Lema\^itre-Robertson-Walker cosmology coupled to a massive scalar field. In particular we show that the focussing of the Liouville measure on attractor solutions is recovered by properly
dealing with a gauge degree of freedom related to the rescaling of the spatial volume. Furthermore, we show how the Liouville measure formulated on a surface of constant Hubble rate, together with the assumption of constant a priory probability, induces a non-uniform probability distribution function on any other surfaces of other Hubble rates. The attractor behaviour is seen through the focussing of this function on a narrow range of physical observables. This qualitative behaviour is robust under change of
potential and underlying measure. One can then conclude that standard techniques from Hamiltonian dynamics suffice to provide a satisfactory description of attractor solutions and the measure problem for inflationary dynamics. 

\end{abstract}

\maketitle

When examining the dynamics of a flat universe whose expansion is driven by a scalar field a curious phenomenon occurs: Dynamical trajectories appear to converge on a set of physical parameters. By a judicious choice of the scalar potential, this convergence can be made to coincide with the observed inflationary phenomena. In particular, when the potential is chosen to be quadratic with mass $m=1.51 \times 10^{-6}$ times the Planck mass, we see that the vast majority of trajectories appear to be focussed on a point consistent with the observations of the spectral index and amplitude of the scalar power spectrum as seen by the Planck and WMAP satellites \cite{Measure2}.\footnote{This mass differs from those previously used as it is based on the latest available data from the Planck satellite \cite{Ade:2013uln}}   

To make quantitative sense of such statements, we require a set of tools adapted to the analysis of inflationary dynamics. In particular, since the space of dynamical trajectories is a continuum, one requires a measure on this space to make sense of any arguments based on counting trajectories. The purpose of this article is therefore twofold: Firstly we will reiterate how such a measure can be canonically constructed on the space of physically distinct solutions, correcting several recent misconceptions about its evolution and relation to thermodynamics. Secondly we will show how the dynamical behaviour of our system explains the attractor behaviour by inducing a probability density function which becomes focussed on points of agreement with observations.

For clarity of exposition, we shall refer throughout this article to flat, homogeneous, isotropic solutions to General Relativity (GR) coupled to a scalar field $\phi$. Though our method applies more broadly, we shall consider the evolution of a single field subject to a quadratic potential $m^2\phi^2/2$. The governing equations of motion are the Friedman and Klein-Gordon equations:
\ba \label{EOM} H^2 &=& \f{4 \pi G}{3} (\dot{\phi}^2 + m^2 \phi^2) \nonumber \\
 \ddot{\phi}&+&3H\dot{\phi}+m^2\phi=0 \ea
In these equations $H$ is the Hubble rate and $m$ the mass of the inflaton $\phi$.  The equations are governed by a Hamiltonian and symplectic structure:
\ba \H= -\f{3vH^2}{8\pi} +  \f{v m^2 \phi^2}{2} +\f{P_\phi^2}{2v} \nonumber \\
    \omega = \d v \wedge \d H + \d\phi \wedge \d P_\phi\, , \ea
where $v$ is the volume and $P_\phi=v\dot{\phi}$. Note that there is a gauge freedom here under rescaling $v$ and $P_\phi$, which keeps the physical degrees of freedom fixed \cite{Measure2,DaveThesis}. Specifically there exists a gauge group, $\mathcal{G}$ which acts on the phase space by transforming
\be (v,H,\phi, P_\phi) \rightarrow (\alpha v, H, \phi, \alpha P_\phi). \ee
This transformation can be generated by a vector field, $\mathbf{g} = v \f{\partial}{\partial v} + P_\phi \f{\partial}{\partial P_\phi}$. This rescaling respects both  the constraint and symplectic structure, so that $\Lie_\mathbf{g} \H = \H$, and $\Lie_\mathbf{g} \omega=\omega$. 

We take two observations from the WMAP and Planck satellites: The spectral index $n_S$ and field amplitude $A$ at the time when a wave of fixed wavelength $k^*$ exited the Hubble radius. These two parameters determine the value of $H$ and $\dot{H}$ at the time set by $k^*$. The equations of motion (\ref{EOM}) relate $H$ to the total energy of the scalar field, and $\dot{H}$ to the kinetic energy. The main question at hand can thus be expressed: What is the probability that the evolution of the system is such that the kinetic energy is compatible with the observations at the time when the total energy is that observed?\footnote{This question is a refinement of that previously formulated in terms of numbers of e-foldings \cite{Gibbons:2006pa}. Although the two are closely related, it is easy to show that a small inflaton mass can give rise to a system with a large number of e-foldings which is not compatible with the satellite data.}.

To answer the question we should count all possible trajectories and find the fraction of them which match the conditions given. To perform such a counting we require a measure on the space of solutions $S$. Such a measure should be i) `Natural'; ii) dependent only on physical degrees of freedom; and iii) count trajectories uniquely. Let us consider these conditions.

The dynamics of our system is such that the Hubble parameter (equivalently energy density) is monotonically decreasing.\footnote{This differs from other approaches that modify GR  such as those used in \cite{Measure, Measure2}, in which quantum effects are included, where the energy density and Hubble rate are only monotonic on sections of solutions.} It is therefore convenient to choose a surface of constant Hubble rate on which to form our measure. This differs qualitatively from the approach described in \cite{Remmen:2013eja} where a measure is sought on the entirety of a phase space reduced to two dimensions. The procedure we use identifies points which are linked by a dynamical trajectory, thus we count each physically distinct solution once only.

The physical degrees of freedom of our system consist of the scalar field, its velocity and the Hubble rate. These are, therefore, the only factors upon which any final result should depend. 

Requiring that a measure be `natural' requires further explanation: Since our measure is to be composed of integrals over physical degrees of freedom, one could introduce any function of these variables into the integrand and construct a new measure, and by judicious choice of function obtain any result. To constrain this we appeal to the `principle of indifference' of Laplace, which states that the distribution which contains least amount of information should be applied. Since our phase space is naturally equipped with a symplectic form - the Liouville measure - this principle leads us to use the uniform distribution on this measure\footnote{Note that this procedure does not in any way invoke thermodynamics. Although the Liouville measure is associated with the entropy of a Hamiltonian system, \textit{the motivation for its use is based solely on its role as symplectic structure}. This renders the criticisms of \cite{Wald} moot since there is no need to appeal to ergodicity or other factors arising from a statistical mechanics viewpoint to justify its use. The measure arises simply from the conditions discussed above.}.
Throughout this paper we will use the Liouville measure, as established by Gibbons, Hawking and Stewart \cite{Gibbons:1986xk}. It is not our aim here to justify the use of such a measure, given that at some level all such choices are ad-hoc \cite{Wald}, but rather to explain the resulting concentration of the measure upon a narrowing range of parameters -the apparent attractor behaviour- and reconcile this with the conservation established in Liouville's theorem. 

Thus the procedure is to pull back the symplectic two-form $\omega = \d v \wedge \d H + \d\phi \wedge \d P_\phi$ onto a surface of constant Hubble rate, following \cite{Gibbons:1986xk,Gibbons:2006pa}. On this surface the first term vanishes as $H$ is constant. We are left with its pullback $\underleftarrow{\omega}$ given by:
\ba \underleftarrow{\omega} &=& \d\phi \wedge \d P_\phi \\
                           &=& \dot{\phi}\, \d\phi \wedge \d v + v\, \d\phi \wedge \d\dot{\phi} \nonumber \\
                           &=& \sqrt{\f{3H^2}{4\pi G} -m^2 \phi^2}\; \d\phi \wedge \d v  \nonumber \ea
where in the second line we have used that $\dot{\phi}=P_\phi/v$. In the final step we note that on constant $H$ surfaces, $\d\phi$ and $\d\dot{\phi}$ are parallel, and used the constraints to rewrite $\dot{\phi}$. 

We can now define the probability of agreement with observed data as the ratio of areas of phase space under this measure:
\be 
P(X)=\frac{\int_A \underleftarrow{\omega}}{\int_S \underleftarrow{\omega}} \, ,
\ee
where $A$ is the area of phase space which leads to having sufficiently low kinetic energy at the observed total energy, and $S$ the total phase space. Here is where the crux of the issue lies: Both numerator and denominator will involve integrals over the gauge degree of freedom $v$. This is a non-compact direction thus the integral is infinite. Since the space of physically distinct solutions is not $S$ but in fact $S/\mathcal{G}$ - dividing out the action of the gauge group - an integral over volume will count the same solution many times. This can be dealt with in a number of ways, the most obvious of which is to place a cutoff $v^*$ on $v$. The integrals cancel for any finite cutoff, yielding a cutoff independent probability given by:
\be 
P(X)= \f{1}{N_H} \int_{A_H(\phi)} \sqrt{\f{3H^2}{4\pi G} -m^2 \phi^2}\; \d\phi \, .\label{prob1}
\ee
Here $A_H(\phi)$ is the range of $\phi$, at a given value of $H$, for which trajectories evolve to the window of values that agree with observations, and the total measure is given by
\be 
N_H = \int_{-\phi_m}^{\phi_m} \sqrt{\f{3H^2}{4\pi G} -m^2 \phi^2}\; \d\phi = \f{3H^2}{8Gm} \, ,
\ee
where $\phi_m=\sqrt{3H^2/4\pi G m^2}$ is the maximum values of $\phi$ compatible with the constraint. Then, if we take this measure and a constant probability distribution at the observed amplitude, we can compute the probability, using (\ref{prob1}), that $n_S$ is within its observational error bars \cite{Ade:2013uln} and find it to be $P(|\phi|) \in [2.800,2.804] = 2.2 \times 10^{-4}$. However,  had we taken the same procedure at $H=1$ and traced trajectories until their energy was compatible with observations, the probability  of agreement with observation would be much larger, namely $(1-10^{-5})$.\footnote{These figures are in agreement with those in \cite{Measure2} with the inflaton mass brought into line with current observations and compatible with the e-foldings observations of \cite{Gibbons:2006pa}} Thus, the probability of agreement grows as we increase the value $H=c$ of the Hubble rate at which the measure is constructed. This points to an apparent contradiction, since our probability is based on the Liouville measure which is preserved under Hamiltonian flow. 

The resolution of this apparent contradiction was reported in \cite{Corichi:2010zp}: Although the trajectories are being squeezed in the $\phi$ direction, they are expanding in the direction of $v$. Thus, although $v$ is pure gauge and thus the probability measure independent of choice of the cutoff $v^*$, the cutoff is not preserved under evolution.  This is how Liouville's theorem is preserved: a contraction in trajectories on physically relevant variable $\phi$ is compensated by an expansion along the conjugate variable $v$, that turns out to be a gauge direction. This also explains how the attractor behaviour arises. 
To summarize, we have seen that the Probability $P$ of having observables compatible with observation is a well defined quantity, given by Eq.~(\ref{prob1}). This quantity though, depends on the choice of $H=c$ surface one takes to compute it. 

Let us now ask a different question. We shall start by considering a uniform probability distribution
at a surface of constant Hubble rate $H_i$, and ask what is the induced {\it probability distribution function} on a different $H_f$ surface that yields {\it the same probability} $P$ for observational compatibility. To do that, we shall first consider how the measure gets modified through the evolution.
To recover such measure on one surface from that on another we must take into account the dependence of the cutoff on the trajectory. In particular, since physically distinct trajectories can be labelled by their value of $\phi$ at a given $H_i$ surface, we find that in order to capture the same physical trajectories on a different, $H_f$ surface, we must let the cutoff be a function $f$ of $\phi$:
\be 
P(X)= \f{1}{N_{(H,\lambda)}} \int_{A_H(\phi)} \int_0^{\lambda f(\phi)} \sqrt{\f{3H^2}{4\pi G} -m^2 \phi^2}\;\, \d v \d\phi \, ,
\ee
where $f(\phi)$ is chosen such that the same trajectories are captured, and $ \lambda \in \mathbb{R}^+$ is our new choice of cutoff - $ \lambda \rightarrow \infty$ covers the entire range of $v$. As before, the total volume is given by 
\be 
N_{(H,\lambda)}= \int_{-\phi_m}^{\phi_m} \int_0^{\lambda f(\phi)} \sqrt{\f{3H^2}{4\pi G} -m^2 \phi^2}\; 
\, \d v \d\phi \, .
\ee
We can trivially perform the integral over volume, to find:
\be  P(X)= \f{1}{N_H} \int_{A_H(\phi)} f(\phi) \sqrt{\f{3H^2}{4\pi G} -m^2 \phi^2}\;\, \d\phi \ee
Thus the cutoff on volume has taken on the role of a probability distribution function (PDF) $f(\phi)$. It is not possible to calculate this function analytically as this would require exact solutions of the inflaton equations of motion. However, numerical integration of the equations can be used to find concrete values, which are shown in the Figure.

\begin{figure}[h!]
\includegraphics[width=0.9\textwidth]{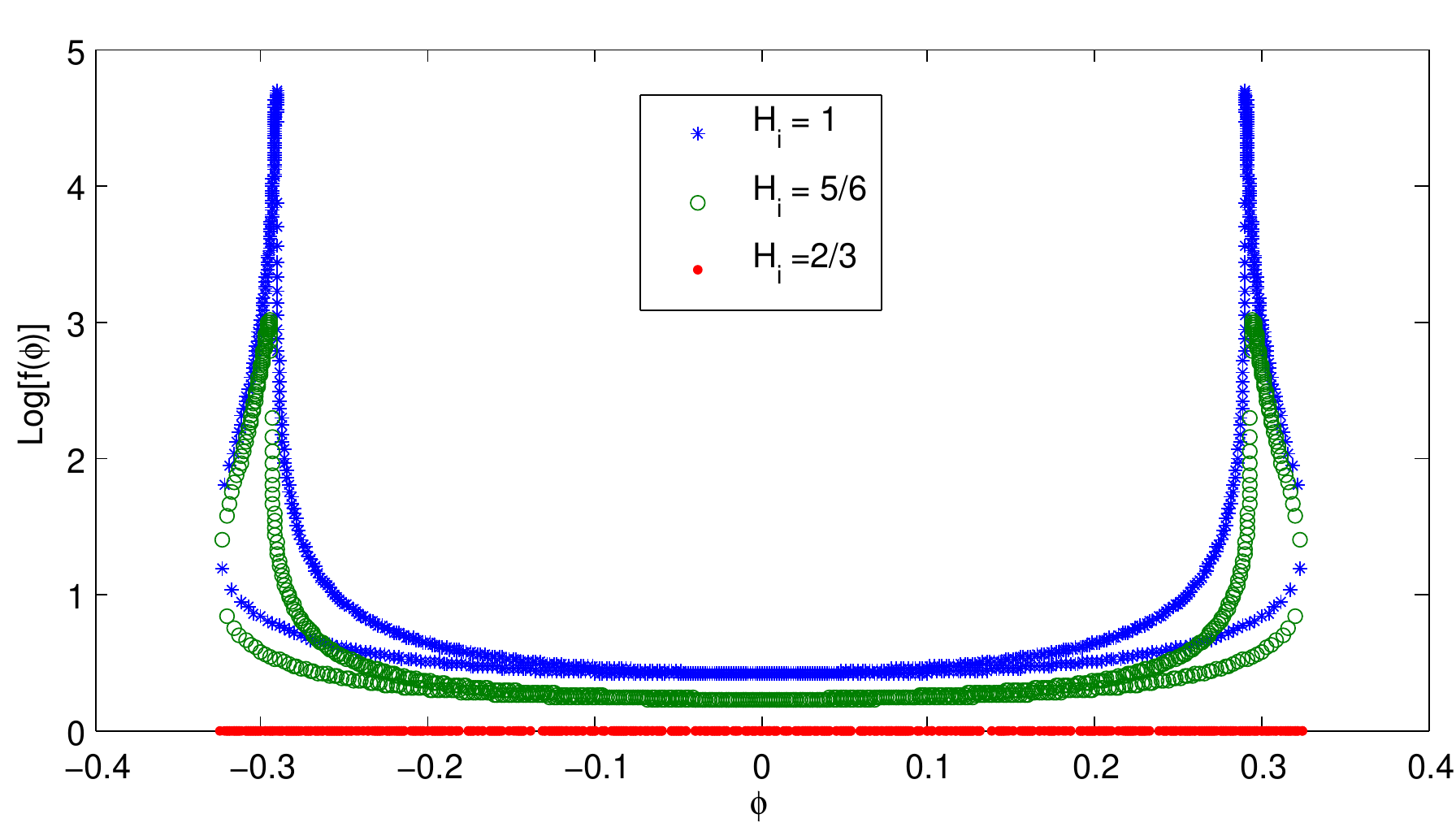} 
\label{fig1}
\caption{The natural logarithms of the probability distribution function $f(\phi)$ induced on a surface of constant Hubble rate from considering the measure on surfaces of higher Hubble rate.  For this figure, the mass of the inflaton was set to $1$, the final surface Hubble rate of $2/3$ (red, dots) and initial surfaces of rate $H=1,5/6$. (blue asterisks, green circles)}
\end{figure}

The function, $f(\phi)$ is established such that the range of values of volume within the cutoff at one Hubble slice is mapped onto a new range at the later slice. Essentially, it can be established in the following way: Since separate trajectories as determined by $\phi$ on an initial Hubble slice expand to different volumes on a later slice, in order to count the same trajectories at a later slice the cutoff of volume should be altered on this later slice. This cutoff dependence on $\phi$ is such that the same set of trajectories captured on the initial slice are measured. This introduces a dependence on $\phi$ in the integral over volume, which can be directly evaluated and thus determines $f(\phi)$. Put succinctly, $f(\phi)$ is therefore determined, up to an overall normalization, as the relative increase in volume of trajectories between slices ending on the final slice with field value $\phi$. 

Here we see that as we increase the initial Hubble rate, the PDF becomes increasingly sharply peaked around certain values. Note that what is plotted in the figure is the logarithm of the PDF - the peakedness rapidly becomes so extreme that it is difficult to display anything other than the value at the peak. The distribution appears double valued, a result of there being two trajectories intersecting each final $\phi$ - one with $\dot{\phi}$ positive, the other negative. For the purposes of illustration, what is plotted is the non-normalized function - normalization would simply add a factor to both numerator and denominator in any probability calculation and thus cancel.

At this point it is important to note that our methods have not depended qualitatively upon the form of the potential used. For more generic potentials, the analysis runs parallel to that presented, and the relative amplitudes of the PDF on final state will be the ratios of the volume expansions that correspond to trajectories evolving to those states. Trajectories with the greatest expansion have the highest induced probability, a naturally induced analogue of volume weighting \cite{Linde:2007nm}. In the case considered here, the massive scalar field, the asymptotic behaviour of the induced PDF is that it becomes a pair of delta functions about the values of $\phi$ compatible with observations. Let us here clarify our statement: The quantitative behaviour and precise predictions of the model (amplitude, spectral index etc.) will depend closely upon the choice of potential. However, the presence of an attractor is purely a feature of the expansion of volume. By a different choice of potential, the location of the attractor can be chosen to take any value of $\phi$ at a given Hubble, but for each potential \textit{an} attractor will exist. 

Let us return briefly to the issue of the choice of measure. As we have stated, any such choice of measure will be ad-hoc, at least in the absence of a more complete quantum theory. However since the volume at a given slice is a gauge choice and the dynamics determine that each trajectory cross a given slice exactly once, any measure can be written in the form 
\be 
\int_A g(\phi)\, l(\phi)\, \d\phi \, ,
\ee
where $A$ is some range of interest in $\phi$ and $l(\phi)$ is the Liouville measure. Since the Liouville measure is conserved, and through conservation asymptotes to a delta function on the values compatible with inflation, the attractor behaviour will be apparent with respect to any measure. One could, of course, choose a measure which highly disfavours inflation on a given slice, and therefore overcome this focussing for some choices of initial Hubble slice. However, as the initial slice is taken to the limit of infinitely large initial Hubble, any given measure will become entirely focussed on those solutions compatible with inflation. To reiterate: the existence of an attractor is independent of the choice of measure, although the choice of measure will determine how such an attractor is approached.

Our analysis to this point has been entirely classical. Eternal inflation, brought about by quantum fluctuations of the scalar field up its potential is believed to occur for sufficiently high values of $\phi$ \cite{Guth:2007ng}. Such fluctuations can cause the total energy of the inflaton to increase, breaking one of the assumptions that went into forming our measure. Therefore, at first glance it appears that our probability calculations would be invalid in this context. However, all eternally inflating trajectories must have a sufficiently high potential energy during the repeating phase. Such trajectories, on exiting the cyclic phase, will all therefore pass through the attractor, since the only phase-space points at such high energy which avoid the attractor are entirely dominated by kinetic energy. That is, the very conditions that are required for eternal inflation to start force the resulting dynamics to pass through the attractor. Since the treatment of dynamics in this paper is entirely classical, we are not in a position to determine the precise nature of eternal inflation. However, since models of eternal inflation typically produce vast numbers of `universes' each of which follows a trajectory following the inflationary model, we can make the following statement: The exit point for eternal inflation is at a sufficiently high value of the Hubble parameter and sufficiently large field value that universes exiting eternal inflation will follow trajectories which pass through the attractor. Whilst we lack the analytical tools to perform a quantitative analysis of eternal inflation itself, the classical trajectories which follow this phase are under control and, subject to the correct potential, will give rise to observations compatible with experimental evidence.  

Let us summarise the main points of this article. For our analysis we have used a natural measure that can be constructed from the symplectic structure and the Liouville measure. As emphasized before, one does not need to rely on thermodynamical arguments to arrive at this description. Such a measure, when evaluated on the space of physically distinct solutions at a fixed Hubble rate, provides a well-defined way to calculate probabilities. The probability 
depends on the choice of constant Hubble surface, assuming a constant probability distribution function.
The attractor behaviour can then be explained within the context of Liouville's theorem --focussing of physical parameters is brought about by the spreading of trajectories along one of the canonical variables--, that turns out to be a gauge direction of phase space. When comparing the probability distributions at two different surfaces,
this expansion was then shown to induce a probability distribution, $f(\phi)$ on a surface of constant $H_f$, starting from another $H_i$ slice, with a constant probability distribution. Furthermore, the attractor behaviour can be seen clearly through the progression of $f$, starting from $f=1$, towards a function heavily weighted towards those solutions compatible with observations. We have argued that this behaviour is robust under the change of potential and initial measure.

As we have shown, a detailed study of the Hamiltonian dynamics and a naturally defined measure, together with an appropriate treatment of a gauge freedom present in the flat FLRW model, is sufficient to explain the apparent existence of attractors, and to provide a qualitative and quantitative explanation of the high probability assigned to phenomenologically favoured physical trajectories.   

\section*{Acknowledgements}

We thank A. Ashtekar and the anonymous referees whose input has greatly improved the paper.
DS is supported by a grant from the Templeton Foundation. This work was in part supported by CONACyT 0177840 and DGAPA-UNAM IN100212 grants.

\bibliographystyle{unsrt}
\bibliography{InflationNotes}

\end{document}